\documentclass[preprint,showpacs]{revtex4}
\usepackage{epsfig}
\newcommand{\eref}[1]{(\ref{#1})}
\newcommand{\Sc}{Schr\"odinger equation }

\newcommand{\si}{\mbox{\boldmath $\sigma$}}
\newcommand{\p}{\mbox{\boldmath $p$}}
\newcommand{\vfi}{\mbox{$\varphi$}}
\newcommand{\sbb}{\mbox{\boldmath $s$}}

\newcommand{\q}{\mbox{\boldmath $q$}}

\newcommand{\hk}{\mbox{\boldmath $\hat{k}$}}
\newcommand{\hr}{\mbox{\boldmath$\hat{r}$}}
\newcommand{\htt}{\mbox{\boldmath$\hat{t}$}}

\newcommand{\I}{\mbox{\boldmath $\rm\bf I$}}
\newcommand{\bT}{\mbox{\boldmath ${\rm\bf T}$}}
\newcommand{\bR}{\mbox{\boldmath ${\rm\bf R}$}}
\newcommand{\bb}{\mbox{${\rm\bf b}$}}
\newcommand{\B}{\mbox{${\rm\bf B}$}}
\newcommand{\hM}{\mbox{\boldmath $\hat{M}$}}
\newcommand{\hX}{\mbox{$\hat{\rm\bf X}$}}

\newcommand{\hT}{\mbox{$\hat{\rm\bf T}$}}
\newcommand{\hR}{\mbox{$\hat{\rm\bf R}$}}
\newcommand\vphi{\varphi}
\newcommand\ola{\overleftarrow}
\newcommand\ora{\overrightarrow}

\newcommand\fr{\displaystyle\frac}
\begin{document}

\title{Reflection of neutrons from fan-like magnetic systems}

\author{V. K. Ignatovich$^1$\footnote{e-mail: ignatovi@nf.jinr.ru}, F. Radu$^2$}

\affiliation{$^1$FLNP JINR Joliot-Curie 6, Dubna, Moscow
reg.141980 RF }\affiliation{ $^2$Helmholtz-Zentrum f\"ur
Materialien und Energie, Albert-Einstein-Str. 15, D-12489, Berlin,
Germany}

\begin{abstract}

An analytical solution is found for neutron reflection
coefficients from magnetic mirrors with fan-like magnetization.
The main feature of the reflection curves related to this type of
magnetization is pointed out. The results of calculations for some
parameters of the system are presented. Time parity and detailed
balance violation in the model are discussed.

\end{abstract}

\maketitle

\section{Introduction}

Investigation of neutron reflection from multilayers belongs to
the field of nanostructure research. Magnetic multilayered systems
with magnetization vector varying in space can be grown
artificially by consecutive evaporation or sputtering techniques
of magnetic layers with different coercivity, or they can appear
naturally under the action of anisotropic exchange
forces~\cite{kort,buz,lei,calv}. A representative  example is a
soft-magnetic layer (SL) which is magnetically coupled to a
hard-magnetic layer (HL)~\cite{goto,dono,kuncser,hellwig}, when
the last one is magnetized non parallel with respect to the
external field orientation. As a result,  the SL becomes virtually
separated into magnetic layers with different orientations of the
magnetization vector, which turns from a direction almost parallel
to the external field to a direction defined by the HL
magnetization vector. This turn can go along a spiral, giving rise
to a helicoidal structure, often referred to as spring magnet, or
this turn can be confined in the plane perpendicular to the sample
interface. In the last case, the magnetization in the SL is
fan-like, and is referred to as perpendicular exchange-spring
multilayer~\cite{perp1,perp2}. Neutron scattering on helicoidal
structures was investigated in Ref.~\cite{aks}. Here we consider
the neutron scattering on a fan-like system. Such a magnetic state
in fact consists of a large number of magnetic layers with
magnetization slightly turned with respect to each other, and the
neutron scattering on them can be calculated numerically by making
use of generalized matrix methods~\cite{rad,polar,ruhm,deak}.
However it is possible to model the fan-like system by a magnet
with continuous rotation of the magnetization vector, and to
calculate neutron scattering on it analytically. We show here that
neutron scattering on fan-like and helicoidal magnets are very
similar. For instance, reflection curves for both configurations
exhibit a resonance in some range of values of the normal wave
vector component of the incident neutrons.

In the next section we find an analytical solution of the
Schr\"odinger equation in a homogeneous medium with fan-like
magnetization. In section III we calculate analytically the
polarized neutron reflectivities from a semi-infinite magnetic
media with fan-like magnetization in the absence of an external
field. In this case, the quantization axis for the neutron spin
$\sbb$ is chosen along the axis $\q$ of the magnetization
rotation, and the neutron scattering becomes dependent on
correlation $\sbb\q$, which violates time parity conservation and
detailed balance theorem.

In section IV we find analytical expressions for reflectivity from
a magnetic slab of finite thickness, also in the absence of an
external field. The results are further used in section V to
calculate the neutron reflection from a magnetic film with
fan-like magnetic state and nonzero external field. In section VI
we consider a SL/HL magnetic heterostructure and calculate the
neutron reflectivity curves from the front side and from the back
side of the sample, respectively like in the~\cite{dono}
experiment. Finally, the results are summarized in section VII.

\section{Neutron propagation in a homogeneous medium with fan-like magnetization}

Let's consider a semi-infinite (z$>$0) mirror and an arbitrary
external field. Magnetic induction in the medium consists of two
parts: one part is constant, $\B=B(1,0,0)$ and directed along an
$x$-axis. The other one, $\bb(z)=b(0,\cos(2qz),\sin(2qz)$, rotates
around $x$-axis (the factor 2 is selected for convenience). Our
task is to find the neutron reflection from such a system.
Therefore, we need to find the neutron wave function inside the
mirror, i.e. to find a solution of the Schr{\"o}dinger equation,
which we shall represent in the form:
\begin{equation}\label{f1}
\bigg(\frac{d^2}{dz^2}-u_0-2b[\sigma_y\cos(2qz+2\varphi)+\sigma_z\sin(2qz+2\varphi)]-
2B\sigma_x+k^2\bigg)|\psi(z)\rangle=0,
\end{equation}
where $\sigma_x$,$\sigma_y$,$\sigma_z$ are the Pauli matrices,
$u_0$ is the medium optical potential multiplied by the factor
$2m/\hbar^2$ ($m$ is the neutron mass), and magnetic fields
include the factor $|\mu|m/\hbar^2$ ($\mu$ is the neutron magnetic
moment).

 In the argument of the rotating field we
included a mismatch phase $2\varphi$, which characterizes the
angle between the external and the internal magnetic fields at the
interface of the medium. In the real class of perpendicular
exchange-spring magnetic multilayers~\cite{perp1,perp2} the
magnetization changes stepwise via magnetic domains with slightly
turned magnetization, but for analytical calculations we can
approximate such a change by a model of a continuous rotation of
this magnetization.

For solution of (\ref{f1}) we use the well known relation:
\begin{equation}\label{hel1}
\sigma_y\cos(2qz+2\vphi)+\sigma_z\sin(2qz+2\vphi)
=\exp(-i\sigma_x(qz+\vphi))\sigma_y\exp(i\sigma_x(qz+\vphi)),
\end{equation}
and represent the wave function $|\psi(z)\rangle$ in the form:
\begin{equation}\label{2}
|\psi(z)\rangle=\exp(-i\sigma_x(qz+\vphi))|\phi(z)\rangle,
\end{equation}
where
\begin{equation}\label{2n}
|\phi(0)\rangle=\exp(i\sigma_x\vphi)|\psi(0)\rangle
\end{equation}
and $|\psi(0)\rangle$ corresponds to an arbitrary spin state.

Substitution of (\ref{hel1}) and (\ref{2}) into Eq.~(\ref{f1})
gives:

\begin{equation}\label{hel2}
\bigg(\frac{d^2}{dz^2}-2iq\sigma_x\frac{d}{dz}-u_0-2b\sigma_y-2B\sigma_x+k^2-q^2\bigg)|\phi(z)\rangle=0.
\end{equation}

\subsection{A solution by Calvo~\cite{calv}}

Such a type of equation was solved in~\cite{calv} by substitution
\begin{equation}\label{calv0}
|\phi(z)\rangle=\exp(ipz)|\xi\rangle
\end{equation}
with constant parameter $p$ and some constant spinor state
$|\xi\rangle$. After substitution we reduce \eref{hel2} to
\begin{equation}\label{calv}
\bigg(-p^2+2q\sigma_xp-u_0-2b\sigma_y-2B\sigma_x+k^2-q^2\bigg)|\xi\rangle=0,
\end{equation}
which is of the type
\begin{equation}\label{calv1}
\hM|\xi\rangle=0
\end{equation}
with constant matrix $\hM$. This equation for nonzero
$|\xi\rangle$ can be satisfied only if det$(\hM)=0$. The last
condition gives an equation for $p$, and for every its root $p_i$
there is the spinor state $|\xi_i\rangle$. Two positive roots give
two spinor eigenstates going along $z$-axis, and two negative
roots give the eigen states doing in opposite directions. It can
be elucidated by analogy with the neutron propagation in a
homogeneous magnetic field $\B$ directed, say, along $x$-axis. The
\Sc equation looks:
$(d^2/dz^2-u_0-2B\sigma_x+k^2)|\phi(z)\rangle=0$. Substitution of
\eref{calv0} reduces it to
$(-p^2-u_0-2B\sigma_x+k^2)|\xi\rangle=0$, which of the form
\eref{calv1}, and the condition det$(\hM)=0$ gives the equation
$(p^2-k^2+u_0)^2-4B^2=0$, from which we get 4 roots
$p_i=\pm\sqrt{k^2-u_0\pm2B}$ for 2 eigenstates $|\xi_{\pm
x}\rangle$ polarized along and opposite $x$-axis.

It is well for description of the neutron propagation in a
homogeneous space. The difficulties start, when we have to
calculate neutron reflection from a magnetic mirror. In that case,
even at a single interface matching of the wave function and its
derivative gives in general 4 linear equations with 4 unknowns,
and if we have two interfaces the number of equations doubles.
Below we show how to avoid this huge number of equations and find
an analytical solution for a general problem of the neutron
reflection and transmission at a magnetic mirror of a finite
thickness.

\subsection{Our solution}

Our solution will proceed along the lines of the work~\cite{aks}.
We will substitute in Eq.~\eref{hel2} a solution for a wave, going
to the right, in the form
\begin{equation}\label{hell0}
|\phi(z)\rangle=\exp(i[a+\ora{\p}\si]z)|\chi\rangle,
\end{equation}
where $|\chi\rangle$ is an arbitrary spin state. This equation has
four unknown parameters $a$ and $\ora{\p}$. A solution for a wave
going to the left can be written as:
\begin{equation}\label{hell0n}
|\phi(z)\rangle=\exp(-i[a+\ola{\p}\si]z)|\chi\rangle
\end{equation}
containing again  four unknown parameters $a$ and $\ola{\p}$.
Here,  $\si=(\sigma_x,\sigma_y,\sigma_z)$ is a vector of the Pauli
matrices.

Substitution of (\ref{hell0}) into (\ref{hel2}) gives:
\begin{equation}\label{hel3}
[-a^2-\ora{\p}^2-2a\ora{\p}\si+2q\sigma_x(a+\ora{\p}\si)
-u_0-2b\sigma_y-2B\sigma_x+k^2-q^2]|\phi(z)\rangle=0.
\end{equation}
The new equation is of the type \eref{calv1}
\begin{equation}\label{alv1}
\hM|\phi(z)\rangle=0.
\end{equation}
It contains a constant matrix $\hM$, but instead of a constant
spinor $|\xi\rangle$ it contains $|\phi(z)\rangle$, which
according to \eref{hell0n} depends on $z$, contains an arbitrary
spinor $|\chi\rangle$, and therefore is itself an arbitrary
spinor. In such a case Eq.~\eref{alv1} can be satisfied only if
$\hM=0$. This condition is equivalent to the following four
equations:
\begin{equation}\label{hhel}
-a^2-\ora{\p}^2+2q\ora{p}_x-u_0+k^2-q^2=0,
\end{equation}
\begin{equation}\label{hel4}
-2a\ora{p}_x+2qa-2B=0,\quad-2a\ora{p}_y-2iq\ora{p}_z-2b=0, \\
\quad -2a\ora{p}_z+2iq\ora{p}_y=0.
\end{equation}
It follows from the last three equations that:
\begin{equation}\label{hel5}
 \ora{p}_x=q-\fr Ba,\quad
\ora{p}_y=\frac{ab}{q^2-a^2},\quad
\ora{p}_z=i\frac{qb}{q^2-a^2}, \\
\quad\ora{\p}^2=\left(q-\fr Ba\right)^2+\frac{b^2}{a^2-q^2}.
\end{equation}
Substitution of (\ref{hel5}) into (\ref{hhel}) leads to:
\begin{equation}\label{hel6}
-a^2+q^2-\fr{B^2}{a^2}-\frac{b^2}{a^2-q^2}-u_0+k^2-q^2=0,
\end{equation}
which is the cubic equation for $a^2$.

\subparagraph{Solution of the cubic equation}

Denoting $x=a^2-q^2$ and $K^2=k^2-u_0-q^2$, we reduce (\ref{hel6})
to
\begin{equation}\label{cu1}
x^3-x^2(K^2-q^2)+x(B^2+b^2-K^2q^2)+b^2q^2=0.
\end{equation}
A further change of variables $x=y+(K^2-q^2)/3$ transforms it to:
\begin{equation}\label{cu2}
y^3-\alpha y-\beta=0,
\end{equation}
where
\begin{equation}\label{cu3}
\alpha=\fr{(K^2-q^2)^2}{3}-B^2-b^2+K^2q^2,\qquad
\beta=\fr{K^2-q^2}{3}\left(\alpha-\fr{(K^2-q^2)^2}{9}\right)+b^2q^2.
\end{equation}
Substituting $y=z+\alpha/3z$, reduces the Eq. (\ref{cu2}) to
$z^3+\alpha^3/27z^3-\beta=0$, and solution of it can be written
as:
\begin{equation}\label{cu4}
z_n=e^{2\pi in/3}\sqrt{\fr{\beta\pm\sqrt{\beta^2-4\alpha^3/27}}2},
\end{equation}
where $n=0,1,2$. Therefore, the function $a(k)$ can be represented
as:
\begin{equation}\label{cu5}
a=\sqrt{q^2+\fr{K^2-q^2}3+z_n+\fr\alpha{3z_n}},
\end{equation}
and the challenge is how to choose a single correct and physical
root. We can look for the correct one between the roots of the
form:
\begin{equation}\label{cu5a}
a=\sqrt{q^2+\fr{K^2-q^2}3+e^{2\pi
in/3}\sqrt[3]{\fr{\beta-\sqrt{\beta^2-4\alpha^3/27}}{2}}+ e^{-2\pi
in/3}\fr\alpha3\sqrt[3]{\fr{2}{\beta-\sqrt{\beta^2-4\alpha^3/27}}}},
\end{equation}
or equivalent ones:
\begin{equation}\label{cu6}
a=\sqrt{q^2+\fr{K^2-q^2}3+e^{2\pi
in/3}\sqrt[3]{\fr{\beta-\sqrt{\beta^2-4\alpha^3/27}}{2}}+ e^{-2\pi
in/3}\sqrt[3]{\fr{\beta+\sqrt{\beta^2-4\alpha^3/27}}{2}}}.
\end{equation}
It is interesting, however, that at some energy interval these two
expressions are not equivalent and (\ref{cu6}) gives erroneous
result, which will be discussed later.

\subparagraph{The choice of roots}

We need to choose such a solution which for small $b$, $B$ and $q$
is reduced to $a=\sqrt{k^2-u_0}$. To make a correct choice, we set
$B=0$ and simplify Eq. (\ref{hel6}) to the quadratic one:
\begin{equation}\label{hel6n}
-a_s^2+q^2-\frac{b^2}{a_s^2-q^2}-K^2=0,
\end{equation}
where $a_s$ denotes $a(B=0)$. The solutions of (\ref{hel6n})
satisfying the above requirement reads:
\begin{equation}\label{llhel7}
a_s=\sqrt{q^2+\frac14\bigg(\sqrt{K^2-2b}+\sqrt{K^2+2b}\bigg)^2}.
\end{equation}

Now, with the help of computer we choose the integer $n$ in the
phase $2\pi n/3$ of (\ref{cu5a}), or (\ref{cu6}), which for $B=0$
gives $a(B=0)$ identical to (\ref{llhel7}). By trial and error
method, we found that in the interval $0<k<k_1$, where $k_1$ has a
value defined by Eq. $Re(\beta^2-4\alpha^3/27)=0$, the integer $n$
is 2 and both forms (\ref{cu5a}), (\ref{cu6}) for the function
$a(k)$ are equivalent; in the interval $k_1<k<k_2$, where
$k_2=\sqrt{u_0+q^2+b}$, $n=1$ and $a(k)$ must be taken in the form
(\ref{cu5a}); above $k_2$ we found $n=0$ and both forms for the
function $a(k)$  are again equivalent. It was found that in the
interval $k_1<k<k_2$ expression (\ref{cu6}) can not be adjusted to
$a_s$, because the computer gives the last term under the first
square root in complex conjugate form comparing to (\ref{cu5a}).

Substitution of (\ref{hell0n}) into (\ref{hel2}) gives:
\begin{equation}\label{hel3a}
[-a^2-\ola{\p}^2-2a\ola{\p}\si-2q\sigma_x(a+\ola{\p}\si)
-u_0-2b\sigma_y-2B\sigma_x+k^2-q^2]|\phi(z)\rangle=0.
\end{equation}
which differs from (\ref{hel3}) only by the sign of $q$. The
relation (\ref{hel3a}) is equivalent to 4 equations similar to
(\ref{hhel}-\ref{hel4}) with negative $q$. Their solutions read:
\begin{equation}\label{hel5extra}
\ola{p}_x=-q-\fr Ba,\quad \ola{p}_y=\frac{ab}{q^2-a^2},\quad
\ola{p}_z=i\frac{-qb}{q^2-a^2},\quad\ola{\p}^2=\left(q+\fr
Ba\right)^2+\frac{b^2}{a^2-q^2},
\end{equation}
and $a$ is identical to (\ref{cu5a}).

\section{Reflection from the interface of a semi infinite fan-like mirror}

In order to calculate reflection from a semi-infinite fan-like
medium we need to match the inside and outside wave functions at
the interface $z=0$. The total wave function for the incident wave
going from the left (in vacuum) toward the interface is equal to:
$$\psi(z)=\Theta(z<0)\left(\exp(i\hk_0z)+\exp(-i\hk_0z)\hr\right)|\xi_0\rangle+$$
\begin{equation}\label{hel81}
+\Theta(z>0)\exp(-i\sigma_xqz)\exp(i[a+\ora{\p}_\varphi\si]z)\htt|\xi_0\rangle,
\end{equation}
where $\Theta(z)$ is a step function, which is equal to unity when
inequality in its argument is satisfied, and to zero in the
opposite case, and
\begin{equation}\label{fan1}
\ora{\p}_\varphi\si=\exp(-i\sigma_x\vphi)\ora{\p}\si\exp(i\sigma_x\vphi).
\end{equation}
The function (\ref{hel81}) at $z<0$ contains an incident wave with
an arbitrary spin state $|\xi_0\rangle$ and a reflected one with
the matrix reflection amplitude $\hr$. The wave vector in the
external field $\B_0$ is the matrix $\hk_0=\sqrt{k^2-2\B_0\si}$.

The internal function is represented by (\ref{2}) with account of
(\ref{2n}) and (\ref{hell0}), where
$|\chi\rangle=|\phi(0)\rangle$, and we replaced $|\psi(0)\rangle$
with $\htt|\xi_0\rangle$, where $\htt$ is the transmission
(refraction) matrix amplitude. Requirement of continuity for the
function (\ref{hel81}) and its derivative at the point $z=0$ leads
to the following equations~\cite{rad,aks}:
$$1+\hr=\htt,$$
\begin{equation}\label{hel0}
\hk_0[1-\hr]=[a-q\sigma_x+
\ora{\p}_\vphi\si]\htt=(a+\ora{\p}'_\vphi\si)\htt,
\end{equation}
where
\begin{equation}\label{hel11}
\overrightarrow{\p}'_\vphi\si=e^{-i\sigma_x\vphi}(\ora{p}_y\sigma_y+\ora{p}_z\sigma_z)
e^{i\sigma_x\vphi}-\sigma_xB/a.
\end{equation}
The solution of the equations (\ref{hel0}) is:
\begin{equation}\label{hela12}
\htt=(\hk_0+a+\ora{\p}'_\vphi\si)^{-1}2\hk_0,
\end{equation}
\begin{equation}\label{hel12}
\hr=(\hk_0+a+\ora{\p}'_\vphi\si)^{-1}(\hk_0-a-\ora{\p}'_\vphi\si).
\end{equation}
It is easy to verify  that at $b=B=0$ one obtains:
\begin{equation}\label{hel13}
\htt=\frac{2\hk_0}{\hk_0+k'},\quad
\hr=\frac{\hk_0-k'}{\hk_0+k'},\quad k'=\sqrt{k^2-u_0},
\end{equation}
which is natural, because at $b=B=0$ the rotation of the
magnetization vector in medium is absent. On the other side, in
the limit $q=0$ we obtain the  formulas for refraction at the
interface of a medium with a uniform magnetization $\bb+\B$.

\begin{figure}[!t]
\centerline{\includegraphics[width=5.60in,height=2.31in]{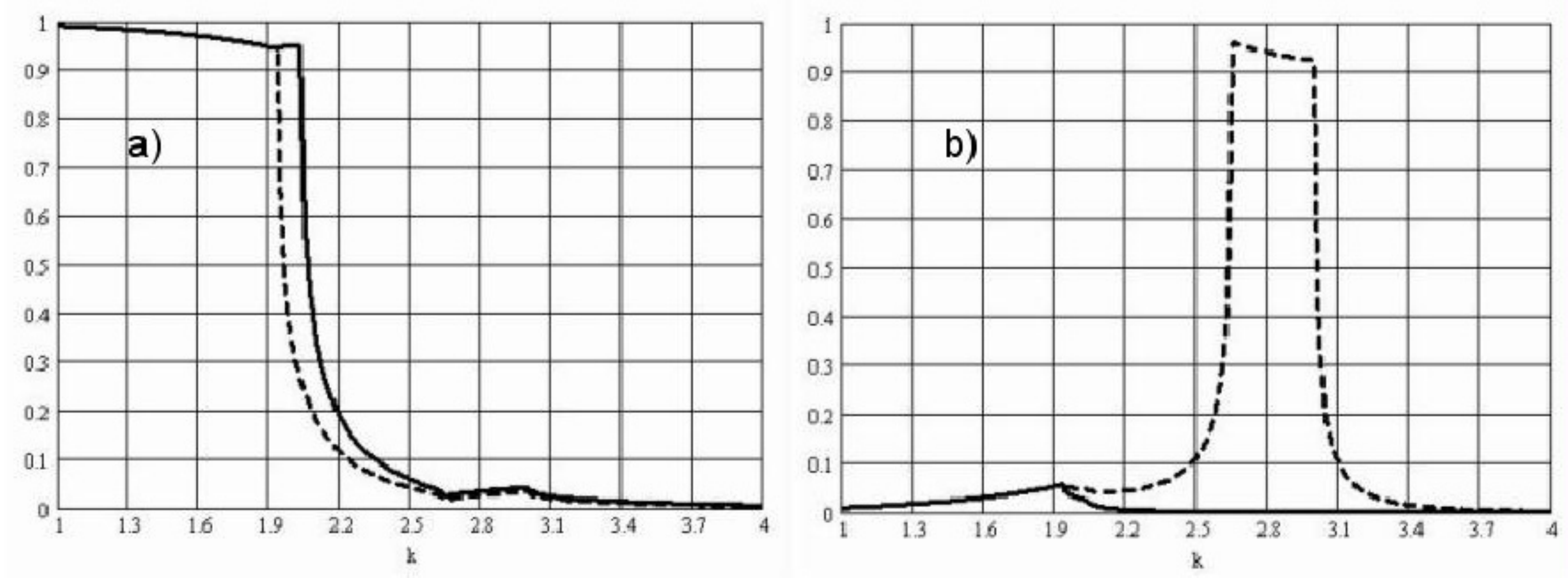}}
\caption{\label{3ff1}Dependence on wave number $k$ of
reflectivities a) without and b) with spin flip. The solid curves
are for initial polarization along $x$-axis, and the dashed curves
are for the initial polarization along $-x$. Calculations were
made for parameters $\B_0=0$, $u_0=4-0.01i$, $b=0.5$, $B=0.1$,
$q=2$.}
\end{figure}

Making use of the analytical expression (\ref{hel12}) we can
easily calculate reflection coefficients from a semiinfinite
mirror with and without spin flip. Their dependence on the wave
number $k$ of the incident neutrons are shown in Fig. \ref{3ff1}
for the simplest case $\B_0=0$, when quantization axis is chosen
along the rotation vector $\q$, i.e. along the $x$-axis. In
calculations, the unity of $k$ is defined via $\sqrt{u_0}/2$. The
optical potential in these units is $u_0=4-0,0001i$ and the other
parameters are $q=2$, $b=0.5$ and $B=0.1$.

The most striking feature of the Fig.~\ref{3ff1}b), is the
presence of a resonant peak with the center at the point
$k=\sqrt{q^2+u_0}$ and the width determined by the field
$\sqrt{b^2+B^2}$. The peak corresponds to almost total reflection
with spin-flip, and it exists only for polarization of the
incident neutron against the vector $\q$, which characterizes the
rotation of the field in the mirror.
\begin{figure}[!b]
\centerline{\includegraphics[width=5.60in,height=2.31in]{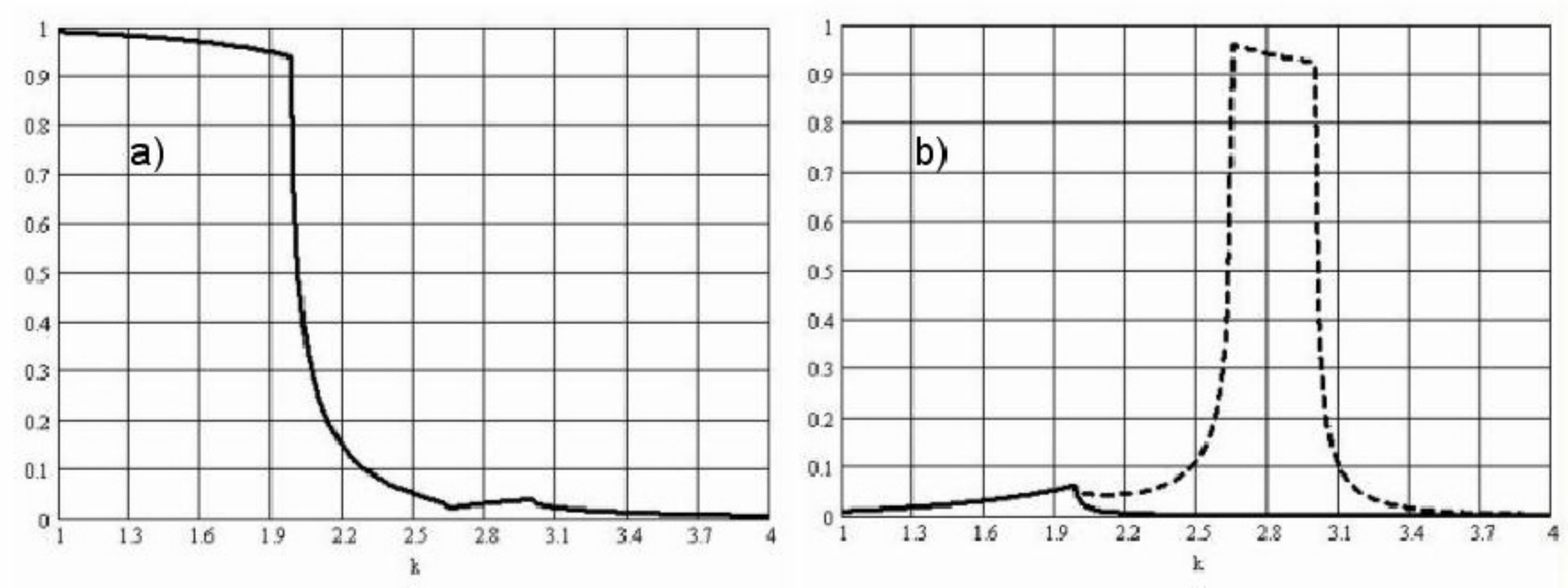}}
\caption{\label{ff1}Dependence on wave number $k$ of
reflectivities a) without and b) with spin flip. The solid curves
are for initial polarization along $x$-axis, and dashed curves are
for initial polarization along $-x$. Calculations were made for
parameters $\B_0=0$, $u_0=4-0.01i$, $b=0.5$, $B=0$, $q=2$. We see
that the center of the peak is at $k=\sqrt{u_0+q^2}$, and almost
total reflection with spin-flip takes place in the range
$\sqrt{u_0+q^2-2b}<k<\sqrt{u_0+q^2-2b}$.}
\end{figure}

It is interesting to compare the obtained results to the case
$B=0$, when we can use the quadratic equation (\ref{hel6n})
instead of the cubic one (\ref{hel6}). The results for the same
parameters except $B=0$ are shown in Fig.~\ref{ff1}. We see that
the main difference is the presence of two limiting energies at
$B\ne0$ for non spin-flip reflectivities. If we are interested in
some peculiarities of reflectivities due to field rotation, we can
neglect $B$, then we get quadratic equation (\ref{hel6n}) instead
of the cubic one (\ref{hel6}), and calculations are facilitated.
So, below we set $B=0$.

\section{Reflection from a mirror of a finite thickness}

To find the reflection coefficients from a finite mirror at
$0<z<l$, we need to know reflection and transmission matrices at
interfaces for a wave incident at the from inside  the mirror. At
the interface $z=0$ the wave function can be written as:
$$|\psi(z)\rangle=\Theta(z<0)\exp(-i\hk_0z)\htt'|\xi_0\rangle+$$
\begin{equation}\label{hel8}
+\Theta(z>0)\exp(-i\sigma_xqz)\left(\exp(-i[a+\ola{\p}_\varphi\si]z)+
\exp(i[a+\ora{\p}_\varphi\si]z)\hr'\right)|\xi(0)\rangle,
\end{equation}
where $\ola{\p}_\varphi\si$ is defined like in (\ref{fan1}).
Matching  this wave function at $z=0$ gives:
\begin{equation}\label{hel16}
\htt'=[\hk_0+a+\overrightarrow{\p}'_\vphi\si]^{-1}[2a+\ora{\p}'_\vphi\si
+\ola{\p}'_\vphi\si], \end{equation}
\begin{equation}\label{hela16}
\hr'=[\hk_0+a+\ora{\p}'_\vphi\si]^{-1}[a+\ola{\p}'_\vphi\si-\hk_0],
\end{equation}
where
$\ola{\p}'_\vphi\si=e^{-i\sigma_x\vphi}(\ola{p}_y\sigma_y+\ola{p}_z\sigma_z)
e^{i\sigma_x\vphi}$.

In order to obtain the reflection and transmission matrices at
$z=l$ it is convenient to shift the origin $z=0$ to this point. As
a result the wave function there becomes:
$$\Theta(z<0)\exp(-iq\sigma_xz)\bigg[\exp(-i[a+\overleftarrow{\p}_\phi\si]z)\hr''+
\exp(i[a+\overrightarrow{\p}_\phi\si]z)\bigg]\xi_0+$$
\begin{equation}\label{hel17}
+\Theta(z>0)\exp(i\hk_0z)\htt''\xi_0,
\end{equation}
and we must take into account that $\htt''\ne\htt'$,
$\hr''\ne\hr'$ and that the mismatch phase $\phi$ is not equal to
$\vphi$ at the entrance surface. Matching  this wave function at
the interface leads to:
\begin{equation}\label{hel18a}
\htt''=[\hk_0+a+\ola{\p}'_\phi\si]^{-1}[2a+\ola{\p}'_\phi\si+\ora{\p}'_\phi\si],
\end{equation}
\begin{equation}\label{hela18a}
\hr''=[\hk_0+a+\ola{\p}'_\phi\si]^{-1}[a+\ora{\p}'_\phi\si-\hk_0].
\end{equation}
At the point $z=-l$ near the entrance surface the reflected wave
of (\ref{hel17}) is equal to
$\exp(iq\sigma_xl)\exp(i[a+\ola{\p}_\phi\si]l)\hr''\xi_0$.

Now we are well equipped to calculate reflection from a slab of
thickness $l$. To facilitate slightly the calculations, we accept
interface $\vphi=0$ at the entrance. Then, at the exit interface
we will have $\vphi=ql$. Let's denote the wave incident from
inside the matter upon the exit interface at $z=l$ as $\hX\xi_0$.
For $\hX$ we can construct a self consistent equation:
$$\hX=\exp(-iq\sigma_xl) \exp(i[a+\ora{\p}\si]l)\htt+$$
\begin{equation}\label{hel20}
+\exp(-iq\sigma_xl) \exp(i[a+\overrightarrow{\p}\si]l)\hr'
\exp(iq\sigma_xl)\exp(i[a+\overleftarrow{\p}_{qL}\si]l)\hr''\hX,
\end{equation}
which has the following solution:
$$\hX=\bigg[I-\exp(-iq\sigma_xl) \exp(i[a+\ora{\p}\si]l)\hr'
\exp(iq\sigma_xl)\exp(i[a+\ola{\p}_{q}\si]L)\hr''\bigg]^{-1}\times$$
\begin{equation}\label{hel21}
\times\exp(-iq\sigma_xl) \exp(i[a+\ora{\p}\si]l)\htt,
\end{equation}
where $I$ is the unit matrix. With the help of $\hX$ we can easily
find the reflection, $\hR$, and the transmission, $\hT$, matrix
amplitudes, which read:
\begin{figure}[!t]
\centerline
{\includegraphics[width=5.60in,height=2.31in]{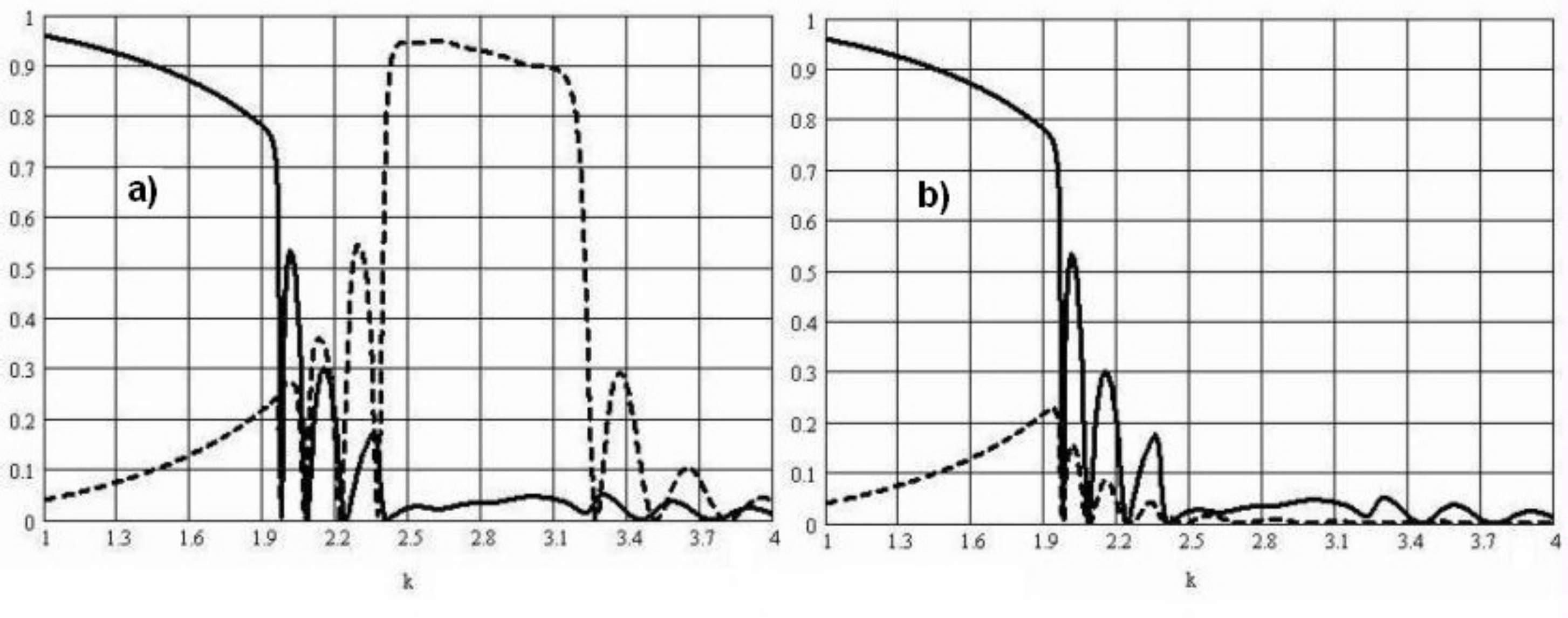}}
\caption{\label{fgig2}Dependence on $k$ of the reflectivities with
(dashed curve) and without (solid curve) of the spin-flip from the
mirror of thickness $l = 8$ with fan-like magnetization. Initial
polarization of the neutron is a) along $-x$, b) along $+x$. For
calculations we used parameters $B_0=0$, $u_0=4-0.01i$, $q=2$, and
$b=1$.}
\end{figure}
\begin{equation}
\label{eq35} {\rm {\bf \vec {R}}} = {\rm {\bf \hat {r}}} + {\rm
{\bf \hat {t}}}'\exp \left( {i\,q\sigma _z l} \right)\exp \left(
{i\,\left[ {a + {\rm {\bf
\mathord{\buildrel{\lower3pt\hbox{$\scriptscriptstyle\leftharpoonup$}}\over
{p}} }}_{qL} {\rm {\bf \sigma }}} \right]\,l} \right)\,\,{\rm {\bf
\hat {r}}}^{''}{\rm {\bf \hat {{\rm X}}}}, \quad {\rm {\bf \hat
{T}}} = {\rm {\bf \hat {t}}}^{''}{\rm {\bf \hat {{\rm X}}}}.
\end{equation}
Using the analytical expressions (\ref{eq35}) we can directly
calculate matrix elements and find reflection and transmission
probabilities with and without spin flip as a functions of $k$.
The results of the calculations for $l=8$ and for the simplest
case $B_0 = 0$ are shown in Fig.~\ref{fgig2}. Again, we clearly
see the occurrence of the resonant peak. Its height decreases with
decrease of $l$.
\begin{figure}[!b]
\centerline{
\includegraphics[clip=true,keepaspectratio=true,width=1\linewidth]{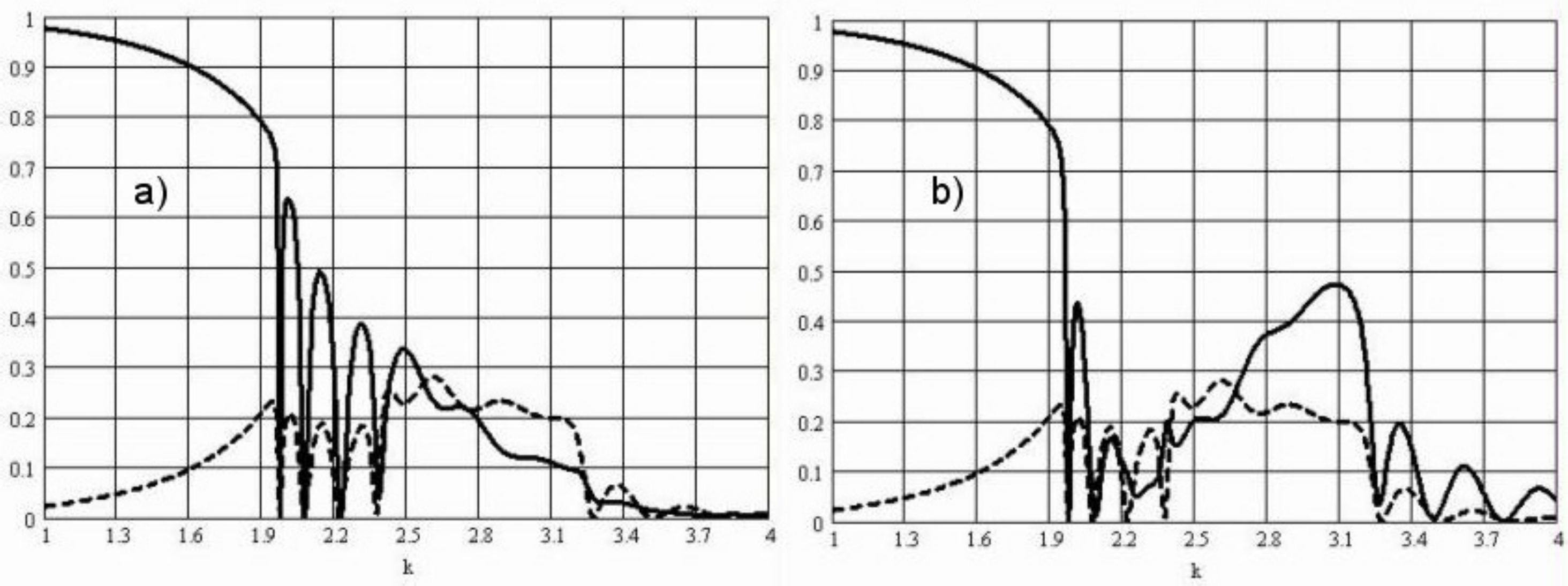}}
\label{fig3}\caption{Dependence on $k$ of the reflectivities with
(dashed curve) and without (solid curve) of the spin-flip from the
mirror with fan-like magnetization in the external field $B_0=0.1$
parallel to the $y$-axis. Initial polarization of the neutron is
a) opposite to $\B_0$, b) along it. Parameters $u_0 $, $q$, $l$
and $b$ are the same as in Fig.~3. The mismatch phase $\varphi$ at
the entrance surface is zero.}
\end{figure}

\section{Reflectivities for $\B_0\ne0$}

When the magnetic induction $\B_0\ne0$, the quantization axis is
to be chosen along $\B_0$. If $\B_0$ is parallel oriented with
respect  to the $y$-axis, then the incident neutron has a
polarization $|\xi_{\pm y}\rangle$, which is a superposition of
two opposite polarizations $|\xi_{\pm x}\rangle$:
\begin{equation}
\label{eq36} \left| {\xi _{\pm y} } \right\rangle = \frac{1\pm
i}{2}\left| {\xi _{ + x} } \right\rangle + \frac{1 \mp i}{2}\left|
{\xi _{ - x} } \right\rangle .
\end{equation}
Therefore, the reflection amplitudes with spin flip, $R_{\pm\mp
y}$, and without spin-flip  $R_{\mp\mp y}$ contain contributions
of the resonant amplitude $R_{+- x}$. This is clearly seen in
Fig.~4. The results of the calculations for a mismatch phase
$\varphi=0$ at the entrance surface show that the amplitudes with
spin-flip are equal for both initial polarizations, and that the
reflection amplitudes without spin-flip also contain a resonant
peak at $\sqrt {u_0 + q^2 - 2b} \le k \le \sqrt {u_0 + q^2 + 2b}
$.

For a real magnetic system~\cite{dono}, the values of the
parameters are about: $q\approx 1$, $ql<\pi$ and $\vfi\ne0$. The
mismatch phase $\vfi$ depends on strength of the field $B_0$ and
can be found from reflectivities curves.

\section{Reflection from composition of soft and hard magnetics}

Above we have considered the reflection from an abstract fan-like
magnetic system. In reality, a fan-like magnetization state can
appear in a SL which is  magnetically coupled to a HL. In the
following we assume that the HL of thickness $l_h$ with optical
potential $u_h$ has an uniform magnetization $\bb$, which
coincides with the magnetization of the SL at the exit interface
$z=l$. The total reflection matrix amplitude $\bR_{sh}$ for the
two magnetic layers and from the side of the soft one is
~\cite{goto,dono}:

\begin{equation}\label{sh}
\bR_{sh}=\ora\bR+\ola\bT\bR_h[\I-\ola\bR\bR_h]^{-1}\ora\bT,
\end{equation}
and from the side of the hard one is:
\begin{equation}\label{hs}
\bR_{hs}=\bR_h+\bT_h\ola\bR_h[\I-\ola\bR\bR_h]^{-1}\bT_h,
\end{equation}
where $\bR_h$ and $\bT_h$ are the reflection and transmission
matrices of the separate HL with vacuum and for the same external
field on both sides of it:

\begin{equation}
\label{eq42} {\rm {\bf R}}_h = {\rm {\bf r}}_h - ({\rm {\bf I}} -
{\rm {\bf r}}_h )\exp (i{\rm {\bf k}}_h l_h ){\rm {\bf r}}_h [\I -
\exp (i{\rm {\bf k}}_h l_h ){\rm {\bf r}}_h \exp (i{\rm {\bf k}}_h
l_h ){\rm {\bf r}}_h ]^{ - 1}\exp (i{\rm {\bf k}}_h l_h )({\rm
{\bf I}} + {\rm {\bf r}}_h ),
\end{equation}
\begin{equation}\label{ht}
\bT_h=({\rm {\bf I}} - {\rm {\bf r}}_h )[\I - \exp (i{\rm {\bf
k}}_h l_h ){\rm {\bf r}}_h \exp (i{\rm {\bf k}}_h l_h ){\rm {\bf
r}}_h ]^{ - 1}\exp (i{\rm {\bf k}}_h l_h )({\rm {\bf I}} + {\rm
{\bf r}}_h ),
\end{equation}
${\rm {\bf k}}_h = \sqrt {k^2 - u_h - 2{\rm {\bf b}}_h {\rm {\bf
\si }}} $, and ${\rm {\bf r}}_h = ({\rm {\bf k}}_h + {\rm {\bf
k}}_0 )^{ - 1}({\rm {\bf k}}_0 - {\rm {\bf k}}_h )$ is the matrix
reflection amplitude of the interface between the HL and vacuum.

The results of the calculations for $u_h=3.5-0.01i$ and $l_h=1.1$
are shown in Fig.~6. The top panels show reflectivities from the
SL side a) without and b) with spin-flip, and the bottom panels
show the analogous reflectivities from the HL side. The top and
bottom figures are different. The resonance peak seen on the top
figures, is replaced with the second total reflection edge when
reflection is from the HL side. This is in good agreement with
measured reflection curves~\cite{dono}. When the external field
changes, the shapes of the top and bottom curves also change.

The results obtained by analytical methods were checked by
numerical simulations with the generalized matrix
method~\cite{rad,polar}. Both types of calculations are in good
agreement. For the numerical calculation, the SL was subdivided
into 50 sub-layers with different directions of magnetization. The
angular increment between two neighboring sub-layers was constant.

By making use of the numerical method, it is also possible to
calculate reflectivities when rotation of the field is not
uniform, and we can expect that nonuniform rotation will lead to
broadening of the resonant peak and to a lowering of its height.

\begin{figure}[!t]
\centerline{
\includegraphics[clip=true,keepaspectratio=true,width=1\linewidth]{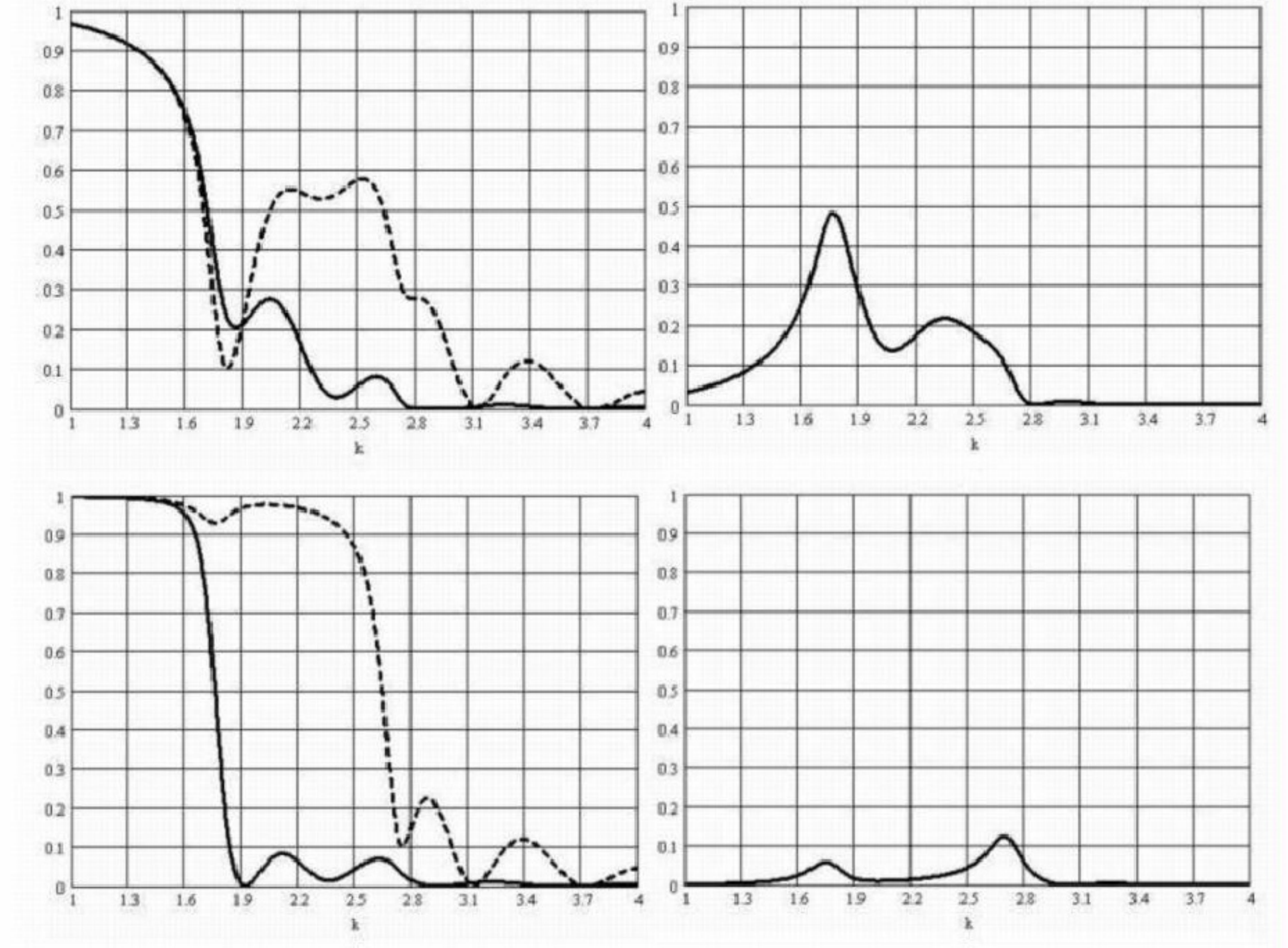}}
\label{fig5}\caption{Dependence on $k$ of the reflectivities from
(top) the SL side (bottom) from HL side. On the left panels the
reflectivities without spin flip when initial polarization is
against field $\B_0$ (solid curve) and along it (dashed curve) are
shown. On the right panels the spin-flip reflectivities are shown.
They are the same for both polarizations, therefore only one curve
is plotted. For the  calculations the following parameters have
been used: $B_0 = 0.2$, $\varphi = 0.1$, $u_0 = 4 - 0.0001i$, $b =
1$, $q = 1$, $l = 3$, $u_h = 3.5 - 0.001i$ and $l_h = 1.1$.}
\end{figure}

\section{Violation of some fundamental principles in the considered model}

The resonant spin flip reflection from samples with rotating field
(see also~\cite{aks})  shows some features, which are very
important with respect to fundamental principles. Reflectivity and
transmissivity of samples with a field rotating around a $\q$
vector contain a T-odd correlation $\sbb\q$, where $\sbb$ is the
incident neutron spin polarization. It looks as if there is a
violation of the time reverse invariance. However more detailed
considerations~\cite{ifn} show that the operator of the time
reverse applied to the spinor wave function, containing T-odd
terms, gives a new wave function, which is an exact solution of
the time reversed Hamiltonian even in presence of absorption
(imaginary part of the Hamiltonian). It proves that the
T-invariance is not violated notwithstanding that the wave
function contains the T-odd correlation. This example, however is
very useful. It shows that observation in an experiment of T-odd
or P-odd correlations should not be accepted as a proof of
violation of P- or T-invariance.

In our problem we meet also violation of the detailed balance
principle, related to the principle of maximal entropy. According
to this principle the flux reflected with spin flip from the
initial state $|\xi_{-x}\rangle$ is to be equal to the flux
reflected with spin flip from the initial state
$|\xi_{+x}\rangle$. Such equality is necessary because, if the
mirror would be immersed in an isotropic distribution of
unpolarized neutrons, the reflectivity would neither change the
isotropy nor create polarization. In the case of the mirror with a
rotating field it is not so. The neutrons in the state
$|\xi_{-x}\rangle$ are reflected almost totally with spin flip,
while the oppositely polarized neutrons are not reflected. It
seems that, if we take a vessel with isotropic unpolarized neutron
gas and partition it off by a magnetized foil with a rotating
magnetization (helicoidal, fan-like or some other) then the
neutrons from, say, the left part of the vessel will spill
completely polarized through the foil into the right part of the
vessel. Indeed the neutrons in the left part, which are polarized
along $-x$, after reflection become polarized along $+x$ and these
neutrons are easily transmitted without depolarization into the
right side. So all the neutrons will go to the right side and they
will be polarized along $+x$.

However such a terrible violation of the maximal entropy principle
does not happen, because for neutrons on the right side of the
foil the field in the foil looks rotating in the opposite
direction, therefore
 the states $|\xi_{+x}\rangle$ are reflected almost totally
with spin flip, and the states $|\xi_{-x}\rangle$ are easily
transmitted to the left side. Nevertheless there appears a cycle
in the phase space, which violates principle of the detailed
balance. This violation is attributed to presence of rotation in
space. If there were two opposite rotations the space could be
considered having an equilibrium state, and the detailed balance
would be not violated, With a single hand rotation the space is
not in the equilibrium, and because of that the detailed balance
is violated.

\section{Conclusion}

We have obtained analytical expressions for calculation of
reflectivities and transmissivity of neutrons through magnetic
media  with fan-like and helicoidal magnetization. The
calculations show that at wave numbers in the interval $\sqrt {u_0
+ q^2 - 2b} < k < \sqrt {u_0 + q^2 + 2b} $, where $q$ is the fan
vector (rotation vector of the fan), and $b$ is the internal
magnetic field of the fan, the reflection curve contains a
resonance peak. This feature serves as an identification of
magnetic configurations with intrinsic rotating fields. The width
of the peak characterizes the uniformity of rotation and the
strength of the rotating field. The  analytical solutions and the
numerical generalized matrix method are very useful tools for
analyzing polarized neutron reflectivity data from  non-collinear
magnets exhibiting anisotropic exchange forces.

The time parity violation found in this model shows that
observation of such an effect in some experiment must be analyzed
in order to exclude the presence of  magnetic fields, which may
rotate in the sample space.

\acknowledgments

We would like to thank Yu. V. Nikitenko for interest, discussions
and support.

\end{document}